# Parsimonious Universal Function Approximator for Elastic and Elasto-Plastic Cavity Expansion Problems

Xiao-Xuan CHEN

PhD student, Department of Civil and Environmental Engineering, The Hong Kong Polytechnic University, Hong Kong, China

Pin ZHANG

Royal Society-Newton International Fellow, Department of Engineering, University of Cambridge, Cambridge, United Kingdom

Corresponding author: dr.pin.zhang@gmail.com

Hai-Sui YU

Professor, School of Civil Engineering, Faculty of Engineering, University of Leeds, United Kingdom

Zhen-Yu YIN

Professor, Department of Civil and Environmental Engineering, The Hong Kong Polytechnic University, Hong Kong, China;

Brian SHEIL

Laing O'Rourke Associate Professor, Department of Engineering, University of Cambridge, Cambridge, United Kingdom

# Parsimonious Universal Function Approximator for Elastic and Elasto-Plastic Cavity Expansion Problems

Xiao-Xuan CHEN, Pin ZHANG[*], Hai-Sui YU, Zhen-Yu Yin and Brian SHEIL

**Abstract:** Cavity expansion is a canonical problem in geotechnics, which can be described by partial differential equations (PDEs) and ordinary differential equations (ODEs). This study explores the potential of using a new solver, a physics-informed neural network (PINN), to calculate the stress field in an expanded cavity in the elastic and elasto-plastic regimes. Whilst PINNs have emerged as an effective universal function approximator for deriving the solutions of a wide range of governing PDEs/ODEs, their ability to solve elasto-plastic problems remains uncertain. A novel parsimonious loss function is first proposed to balance the simplicity and accuracy of PINN. The proposed method is applied to diverse material behaviours in the cavity expansion problem including isotropic, anisotropic elastic media, and elastic-perfectly plastic media with Tresca and Mohr-Coulomb yield criteria. The results indicate that the use of a parsimonious prior information-based loss function is highly beneficial to deriving the approximate solutions of complex PDEs with high accuracy. The present method allows for accurate derivation of solutions for both elastic and plastic mechanical responses of an expanded cavity. It also provides insights into how PINNs can be further advanced to solve more complex problems in geotechnical practice.

# Introduction

Cavity expansion is a canonical problem in geotechnical engineering (Yu and Houlsby, 1991, Yu, 2000). The analysis of the expansion of cylindrical and spherical cavities in soils and rocks has been extensively applied to a wide range of practical engineering problems such as cone penetration tests (Vesić, 1972, Salgado *et al.*, 1997, Diao *et al.*, 2015), pile installation (Randolph *et al.*, 1979, Rezania *et al.*, 2017, Wang *et al.*, 2020), pressuremeter tests (Hughes *et al.*, 1977, Gibson, 1961) and tunnel excavation (Yu and Rowe, 1999, Zhuang *et al.*, 2022). To better capture real-world problems, cavity expansion theory has evolved over the years to incorporate various soil behaviours such as elastoplasticity (Carter *et al.*, 1986, Yu and Houlsby, 1991, Vesić, 1972, Li *et al.*, 2021), state-dependency (Collins *et al.*, 1992, Salgado and Randolph, 2001, Huang *et al.*, 2021), and even more complex phenomena such as particle crushing or size-dependent behaviours of a cavity (Jiang and Sun, 2012, Russell and Khalili, 2002, Liu *et al.*, 2021, Zhuang *et al.*, 2018). Owing to its prominence in the geotechnical literature spanning four decades, cavity expansion has become a widely adopted boundary value problem in its own right.

Cavity expansion analysis is typically initiated by establishing governing partial differential equations (PDEs) or ordinary differential equations (ODEs). These PDEs/ODEs consist of the equilibrium equations, the compatibility equations, the constitutive relations, and the initial and boundary conditions. To solve these PDEs/ODEs, a plethora of analytical, semi-analytical and numerical methods have been proposed (Tang *et al.*, 2021, Chen *et al.*, 2012, Chen and Abousleiman, 2012, Huang *et al.*, 2020, Cheng *et al.*, 2022). Whilst early analytical and semi-analytical solutions were convenient, they were limited to simple cases such as low-dimensional problems and simple constitutive relations (Vesić, 1972, Krishnapriyan *et al.*, 2021). For this reason, numerical methods, such as finite element (Wang *et al.*, 2010, Huang *et al.*, 2004) and finite difference (Zhou *et al.*, 2021), emerged as a more effective means of modelling the cavity expansion process. However, such numerical methods are not without their own drawbacks – computational costs can be prohibitively high, mesh refinement is commonly heuristic and slow, and convergence remains troublesome for complex problems.

Recently, neural networks constrained by prior information (physics/empirical knowledge), which are commonly termed physics-informed neural networks (PINNs) (Raissi *et al.*, 2019), have emerged as a promising method for solving both forward and inverse PDEs/ODEs in a wide range of practical domains, such as fluid mechanics (Cai *et al.*, 2022, Sun *et al.*, 2020, Mao *et al.*, 2020, Jin *et al.*, 2021), heat transfer (Cai *et al.*, 2021, Cai *et al.*, 2020), and solid mechanics (Haghighat *et al.*, 2022, Zhang *et al.*, 2022, Guo and Haghighat, 2022). Compared with conventional numerical methods, PINNs can be considered a mesh-free method that bypasses meshing-related inefficiencies. PINNs are generally accomplished by incorporating physics-based terms (e.g., the initial/boundary conditions and governing PDEs/ODEs) into the loss function of neural networks. The training process aims to minimize the discrepancies between the neural network outputs and the incorporated physics terms, ensuring the final outputs are equivalent to the solutions of the PDEs/ODEs. Unlike conventional machine learning algorithms, PINNs operate as data-free modelling methods. Thus, as long as a comprehensive understanding of physics-based principles is provided and encoded, PINNs can generate the solution of PDEs/ODEs without any labelled data.

Given the advantages of PINNs, recent studies have explored their application to geoengineering problems. For instance, PINNs have recently been applied to Terzaghi's one-dimensional (1D) consolidation to predict pore pressure distributions (forward analysis) and identify the coefficient of consolidation (inverse analysis) with excellent accuracy (Bekele, 2021). PINNs were extended to prior information-based neural networks (PiNets), in the sense that empirical relations were embedded in the neural network loss functions instead of rigorous physical laws; that approach was shown to accurately analyse 1D consolidation problems, considering both surcharge and vacuum loading conditions (Zhang *et al.*, 2023b). The application of physics-informed learning approaches to consolidation problems has also been explored by Mandl *et al.* (2023), and extended to more complex scenarios such as consideration of two dimensions (Lu and Mei, 2022), the nonlinearity of the volume compressibility and hydraulic conductivity coefficients (Lan *et al.*, 2023), and even thermoelastic consolidation of unsaturated strata (Amini *et al.*, 2023). PINNs have also been applied to tunnelling problems. For instance, PINNs have been used for the inverse analysis of the foundation modulus and the proposed framework was subsequently used for the real-world settlement prediction for the Hong Kong–Zhuhai–Macau Bridge tunnel (He *et al.*, 2024b), and this research was further enhanced by considering and identifying the time-dependent characteristics of foundation modulus (He *et al.*, 2024a). Besides, PINNs were adopted for forward and inverse analysis of tunnelling-induced ground elastic deformations (Zhang *et al.*, 2023c). Other geotechnical applications of PINNs/PiNets include modelling unsaturated flow in porous media (Depina *et*

*al.*, 2022, Haghighat *et al.*, 2022, Yang and Mei, 2022), constitutive modelling (Zhang *et al.*, 2023a) and pile-soil interaction modelling (Vahab *et al.*, 2023). Whilst the popularity of PINNs is growing rapidly, their application to geotechnical problems is still in its infancy and there remains a significant gap in applying PINNs in real-world practice. This is largely due to a lack of consideration of more realistic behaviours in the literature such as soil plasticity and multi-physics interactions.

To bridge that gap, this research proposes a novel parsimonious PiNet framework to address the challenge of solving complex partial differential equations (PDEs). The academic contributions of this paper include: (a) introducing a novel parsimonious structure of loss functions which enables accurate solution of complex PDEs using PiNet, and the effectiveness of this parsimony principle is further elucidated through analysing back-propagated gradients of the loss functions, (b) introducing a PiNet framework which is capable of solving forward elasto-plastic geotechnical problems for the first time, while considering both Tresca and Mohr-Coulomb yield criteria to demonstrate model generalisability. The feasibility of this new framework is verified on a canonical spherical cavity expansion problem. The prior knowledge, i.e., equilibrium equations, compatibility equations, constitutive relations and boundary conditions, are integrated into the loss functions of the PiNet. The accuracy and effectiveness of PiNet solutions are validated through comparisons to the analytical solutions (Yu, 2000).

## Methodology

**Overview of the PiNets**

The framework of a PiNet-based PDE solver is illustrated in Fig. 1. In various engineering problems, a general PDE describes the behaviour of a system can be expressed as:

$$u_t(t, \boldsymbol{x}) + \mathcal{D}[u(t, \boldsymbol{x})] = 0, \boldsymbol{x} \in \Omega, t \in [0, T] \tag{1}$$

where $t$ and $\boldsymbol{x}$ are the temporal and spatial coordinates, $u(t, \boldsymbol{x})$ is the unknown solution of this PDE which can be the displacement field or any interested unknown variables, $u_t(t, \boldsymbol{x})$ is the derivative of $u(t, \boldsymbol{x})$ with respect to $t$, $\mathcal{D}[\cdot]$ is a nonlinear differentiation operator, $\Omega$ is the domain of input variables and $T$ is the maximum computation time.

Commonly, the PDE is accompanied by specific conditions at the boundaries of the system, referred to as Dirichlet and Neumann boundary conditions which can be described by:

$$u(t, \boldsymbol{x}) = \mathcal{B}_D(t), \boldsymbol{x} \in \partial\Omega_D, t \in [0, T] \tag{2}$$

$$\mathbf{n} \cdot \nabla u(t, \boldsymbol{x}) = \mathcal{B}_N(t), \boldsymbol{x} \in \partial\Omega_N, t \in [0, T] \tag{3}$$

where $\partial\Omega_D$ and $\partial\Omega_N$ are the Dirichlet and Neumann boundaries, respectively, $\mathcal{B}_D(t)$ and $\mathcal{B}_N(t)$ are the prescribed functions on the Dirichlet and Neumann boundaries, respectively, $\mathbf{n}$ denotes the unit outward normal vector of the boundaries, and $\nabla$ is the gradient operator.

Similarly, the PDE is accompanied by initial conditions that describe the system's behaviour at the beginning of the analysis, providing a starting point for understanding the system's evolution over time:

$$u(0, \boldsymbol{x}) = \mathcal{I}_0(\boldsymbol{x}), \boldsymbol{x} \in \Omega \tag{4}$$

$$u_t(0, \boldsymbol{x}) = \mathcal{I}_1(\boldsymbol{x}), \boldsymbol{x} \in \Omega \tag{5}$$

where $\mathcal{I}_0(x)$ and $\mathcal{I}_1(x)$ are the initial states of the system.

A fully connected deep neural network (DNN) is adopted herein as the baseline algorithm for the PiNet, given that a well-trained DNN with multiple hidden layers can approximate almost any function (Sun *et al.*, 2023). In DNN (Fig. 1), the hidden layers receive the information from the input layer and pass the transformed information to the output layer. This process can be expressed mathematically as:

$$z_0 = x \tag{6}$$

$$z_l = \sigma(\mathbf{W}_l z_{l-1} + b_l), l \in [1, L] \tag{7}$$

$$z_{L+1} = \mathbf{W}_{L+1} z_L + b_{L+1} \tag{8}$$

where $L$ is the total number of the DNN hidden layers, $\sigma$ is the activation function, which is chosen to be *tanh* in this research, $z_{l-1}$ and $z_l$ are the input and output tensors of the $l^{th}$ layer, respectively, $\mathbf{W}_l$ and $b_l$ are the weight matrix and bias vector of the $l^{th}$ layer, respectively, $z_0$ and $z_{L+1}$ are the input and output tensors of the DNN, respectively. The weight $\mathbf{W}$ and bias $b$ are iteratively updated via the gradient-based optimization method such as Adam (Kingma and Ba, 2014) or L-BFGS (Liu and Nocedal, 1989) to minimize the discrepancy between the predictions and exact values.

In the implementation of conventional DNNs, known paired input-output training data are required during training. However, the output $u(t, x)$ of the PiNet is the solution to the PDEs which is not yet known. To achieve an accurate approximation to the solution without the need for direct input-output mapping, prior knowledge of the current problem is incorporated into the loss function of the PiNet. Specifically, the residuals of the governing PDEs/ODEs, initial/boundary conditions and measured data (if available) are embedded into the loss function as constraints. Accordingly, the prior knowledge-embedded loss function can be expressed by the sum of four loss components, consisting of three physics-based terms and one data-based term, as follows:

$$\mathcal{L} = w_1 \mathcal{L}_{PDE} + w_2 \mathcal{L}_{BC} + w_3 \mathcal{L}_{IC} + w_4 \mathcal{L}_{data} \tag{9}$$

where $\mathcal{L}$ is the value of the total loss, $\mathcal{L}_{PDE}$, $\mathcal{L}_{BC}$ and $\mathcal{L}_{IC}$ are the residual values of governing PDEs, boundary conditions and initial conditions, respectively, which are regarded as the incorporated physics constraints in the networks, $\mathcal{L}_{data}$ is the residual value from measured points, which can be regarded as the pure data-driven part in the networks, and $w_1$, $w_2$, $w_3$, $w_4$ are weights of each loss terms which can be used to eliminate the imbalance scale between different loss terms.

The three physics-based terms in the loss function can be expressed as:

$$\mathcal{L}_{PDE} = \frac{1}{n_{col}} \|u_t(t, \boldsymbol{x}) + \mathcal{D}[u(t, \boldsymbol{x})]\|^2 \tag{10}$$

$$\mathcal{L}_{BC} = \frac{1}{n_b} \|u(t, \boldsymbol{x}) - \mathcal{B}_D(t)\|^2 + \frac{1}{n_b} \|\mathbf{n} \cdot \nabla u(t, \boldsymbol{x}) - \mathcal{B}_N(t)\|^2 \tag{11}$$

$$\mathcal{L}_{IC} = \frac{1}{n_i} \|u(0, \boldsymbol{x}) - \mathcal{I}_0(\boldsymbol{x})\|^2 + \frac{1}{n_i} \|u_t(0, \boldsymbol{x}) - \mathcal{I}_1(\boldsymbol{x})\|^2 \tag{12}$$

where $n_{col}$, $n_b$ and $n_i$, are the number of collocation points in the computational domain, the number of sampling points on the boundaries and at initials, respectively, $\frac{1}{n_i}\|\cdot\|^2$ aims to calculate the mean square error (MSE) of the predictions, which is commonly used in most literature to measure the loss (Haghighat *et al.*, 2021, Mao *et al.*, 2020, Raissi *et al.*, 2019).

To incorporate the available measured data into the loss function, the data-based term in the loss function can be expressed as follows:

$$\mathcal{L}_{data} = \frac{1}{n_d}\|u(t,\boldsymbol{x}) - u^*(t,\boldsymbol{x})\|^2 \qquad (13)$$

where $n_d$ is the number of data points, $u$ is the prediction from the PiNet and $u^*$ is the corresponding exact value.

To calculate the differential components in the residual terms, the automatic differentiation (AD) technique is used, which is theoretically based on the chain rule and backpropagation (Baydin *et al.*, 2018). Unlike numerical differentiation techniques such as finite difference, AD does not differentiate data and therefore AD can avoid approximation error and tolerate noisy data (Lu *et al.*, 2021).

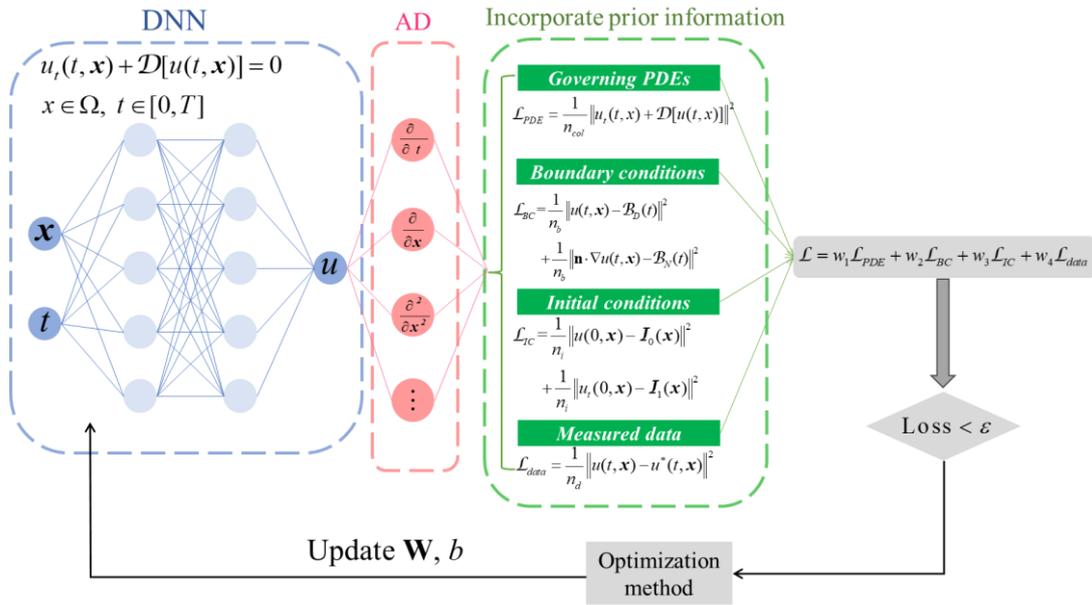

**Fig. 1 Schematic illustration of the proposed PiNet as a PDE solver**

**Governing equations**

In this research, the analytical framework for the governing equations is based on the following assumptions: (1) the soil is considered as a homogenous, isotropic continuum without body forces; (2) the elastic cross-anisotropic soil characterizes transverse isotropy in the radial direction such that the surface normal to the radius is regarded as the isotropic plane; (3) small-strain theory is adopted for cavity expansion modelling and solely focusing on the analysis of stress fields. Based on these assumptions, the whole spherical cavity is axisymmetric and the displacement is radial everywhere. Hence, the spherical cavity expansion problem can be effectively condensed to the 1D form, which also simplifies the formulation of the governing equations into ODEs.

Fig. 2 illustrates the basic definition of the cavity expansion problem. A sphere with inner radius $a$ and outer radius $b$, subjected to an internal pressure $p$ and external pressure $p_0$ is depicted. Compressive stresses and strains are defined as positive. The problem is analysed using the polar coordinate system $(r, \theta)$. When the cavity expansion is relatively small, the soil surrounding the spherical cavity deforms elastically (Fig. 2a). When the cavity expansion is sufficiently large, the soil in the inner layer surrounding the cavity begins to yield, and a plastic zone spreads progressively outwards as the cavity continues to expand (Fig. 2b). The interface between the elastic and plastic region is defined as the elastic-plastic (EP) boundary, and its location is denoted by the plastic zone radius $c$. The principle stresses in the radial and tangential directions are denoted $\sigma_r$ and $\sigma_\theta$, respectively, and the corresponding principle strains are denoted by $\varepsilon_r$ and $\varepsilon_\theta$, respectively.

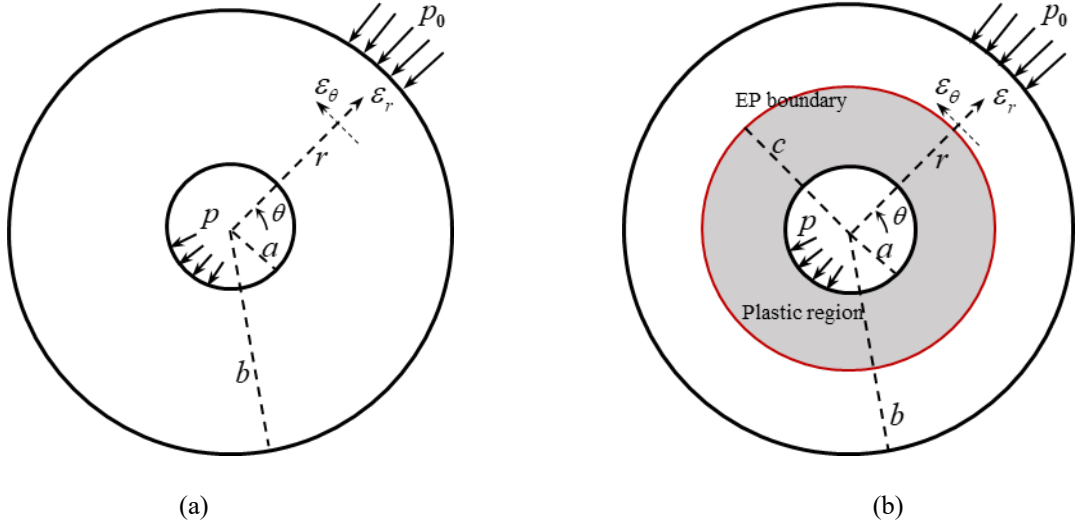

**Fig. 2** Illustration of the cavity expansion problem with: (a) elastic soil behaviour; (b) elasto-plastic soil behaviour

Considering an element at a radial distance $r$ from the centre of the spherical cavity, the equilibrium equation for both elastic and plastic regions is:

$$2(\sigma_r - \sigma_\theta) + r\frac{d\sigma_r}{dr} = 0 \tag{14}$$

*Elastic analysis*

For elastically deformed soils, the normal strains in two directions can be expressed as functions of the radial displacement $u$, which is also known as the compatibility equations:

$$\dot{\varepsilon}_r = -\frac{d\dot{u}}{dr}; \quad \dot{\varepsilon}_\theta = -\frac{\dot{u}}{r} \tag{15}$$

where the symbol ($\dot{\square}$) denotes the incremental value for each variable.

The constitutive relations for elastic isotropic materials and elastic cross-anisotropic materials can be expressed by:

$$\begin{bmatrix} \dot{\varepsilon}_r \\ \dot{\varepsilon}_\theta \end{bmatrix} = \begin{bmatrix} \frac{1}{E} & -\frac{2v}{E} \\ -\frac{v}{E} & \frac{1-v}{E} \end{bmatrix} \begin{bmatrix} \dot{\sigma}_r \\ \dot{\sigma}_\theta \end{bmatrix} \tag{16}$$

$$\begin{bmatrix} \dot{\varepsilon}_r \\ \dot{\varepsilon}_\theta \end{bmatrix} = \begin{bmatrix} \frac{1}{E'} & -\frac{2v}{E} \\ -\frac{v'}{E'} & \frac{1-v}{E} \end{bmatrix} \begin{bmatrix} \dot{\sigma}_r \\ \dot{\sigma}_\theta \end{bmatrix} \tag{17}$$

where $E'$ and $E$ are the Young's moduli in the $r$ direction and in the isotropic surface, respectively, and $v'$ and $v$ are Poisson's ratios in the $r$ direction and in the isotropic plane, respectively.

As the inner and outer pressure are known to be $p$ and $p_0$ in the radial direction, the boundary conditions can be written as:

$$\sigma_r = p \ (r = a); \qquad \sigma_r = p_0 \ (r = b) \tag{18}$$

### *Elasto-plastic analysis*

For the elasto-plastic case, we consider the material to be isotropic in the context of elastic-perfectly plastic constitutive relations. In this case, the soil surrounding the spherical cavity is divided into two regions, an elastic region ($c < r < b$) and a plastic region ($a < r < c$). Obtaining separate elasto-plastic solutions for the elastic and plastic regions requires prior information on the EP boundary defined by the radius of the plastic region $c$ (Liu *et al.*, 2021). The radial and tangential stresses in the elastic and plastic regions are denoted $\sigma_r^e$, $\sigma_\theta^e$ and $\sigma_r^p$, $\sigma_\theta^p$, respectively.

In the elastic region, the equilibrium equation, constitutive relation and compatibility equations are the same as in the elastic isotropic case (Eq. 14-16). The boundary condition on the EP boundary $r = c$ for the elastic region can be obtained according to the yield criterion:

$$\sigma_r - \alpha \sigma_\theta = \sigma_Y \tag{19}$$

where $\sigma_Y$ denotes the soil yield stress, and $\alpha$ is a material parameter. For the Tresca criterion, $\alpha = 1$ and $\sigma_Y = 2s_u$, where $s_u$ is the undrained strength of the soil. For the Mohr-Coulomb criterion, $\alpha = (1 + \sin\phi) / (1 - \sin\phi)$ and $\sigma_Y = 2C\cos\phi / (1 - \sin\phi)$, where $\phi$ and $C$ are the friction angle and cohesion of the soil. These variables are all considered known parameters.

For the plastic region, the equilibrium equation is defined by Eq. 14. For continuity of stress on the EP boundary, the boundary condition on $r = c$ for the plastic region can be obtained from the results of the stress field in the elastic region. Thus, the boundary conditions of the plastic region can be defined as:

$$\sigma_r^p = \sigma_r^e, \ (r = c)$$

$$\sigma_\theta^p = \sigma_\theta^e, \ (r = c) \tag{20}$$

As the entire plastic region is in the yielding state, the yield condition can be satisfied at every point within the plastic region such that Eq. 19 is another governing equation.

## Parsimonious loss function

The parsimony principle, rooted in the pursuit of achieving simplicity and efficiency of the model, holds significant relevance in the context of machine learning and its application to solving complex problems. Since the fundamental principle of physics-informed learning is penalizing loss functions containing known governing equations, the formulation of these governing equations becomes critical. In this study, a parsimony principle is adopted for PiNet training for the first time where the simplified formulations of governing equations are encoded in the PiNet loss functions. Besides, rather than predicting all relevant variables (displacement, strain and stress) simultaneously, the output dimension of the parsimonious PiNet is reduced to only comprise stress components. The motivation behind employing the parsimonious principle is to provide accurate stress predictions for cavity expansion problems while mitigating the computational complexity of PiNets.

Herein, a sensitivity study was performed to explore the influence of different governing equation formulations (embedded in the loss functions) on the PiNet performance, emphasizing the principle of parsimony. By way of example, three different levels of simplification and parsimony for the governing equation embedded loss functions are considered for the elastic isotropic problems (see Table 1).

The most complete formulation (formulation **A**) considers all governing equations directly, i.e., equilibrium equations, constitutive relations, compatibility equations and boundary conditions. This is the most popular approach in the current physics-informed learning research (Haghighat *et al.*, 2021, Zhang *et al.*, 2023c, Abueidda *et al.*, 2021). In this case, the output layer of PiNet contains five variables: $\sigma_r$, $\sigma_\theta$, $\dot{\varepsilon}_r$, $\dot{\varepsilon}_\theta$ and $\dot{u}$.

Formulation **B** combines two compatibility equations (Eq. 15) by eliminating the displacement variable $u$. The simplified compatibility equation is now:

$$\dot{\varepsilon}_r = \frac{\mathrm{d}}{\mathrm{d}r}(r\dot{\varepsilon}_\theta) \tag{21}$$

In this context, the PiNet has four output variables: $\sigma_r$, $\sigma_\theta$, $\dot{\varepsilon}_r$ and $\dot{\varepsilon}_\theta$.

Formulation **C** integrates all compatibility equations and constitutive relations, leading to a significant reduction in the complexity of the loss function and output dimension. Moreover, the strains can also be removed from the output through the combination of compatibility equations and constitutive relations. This procedure further reinforces the parsimony principle with a new mathematical expression of governing equations, which describes the correlation between the stress and radius as:

$$\sigma_r - \sigma_\theta = r\frac{d\sigma_\theta}{dr} \tag{22}$$

As a result, the output variables of the PiNet in this case comprise only two stress components.

Table 1 Three formulations of the governing equations with different levels of complexity

| Different formulations | Output variables | Governing equations |
|---|---|---|
| Formulation A | $\sigma_r, \sigma_\theta, \dot{\varepsilon}_r, \dot{\varepsilon}_\theta, \dot{u}$ | $2(\sigma_r - \sigma_\theta) + r\frac{d\sigma_r}{dr} = 0$ <br> $\dot{\varepsilon}_r = -\frac{d\dot{u}}{dr}; \dot{\varepsilon}_\theta = -\frac{\dot{u}}{r}$ <br> $\dot{\varepsilon}_r = \frac{\dot{\sigma}_r - 2v\dot{\sigma}_\theta}{E}$ <br> $\dot{\varepsilon}_\theta = \frac{-v\dot{\sigma}_r + (1-v)\dot{\sigma}_\theta}{E}$ <br> $\sigma_r = p \ (r=a); \sigma_r = p_0 \ (r=b)$ |
| Formulation B | $\sigma_r, \sigma_\theta, \dot{\varepsilon}_r, \dot{\varepsilon}_\theta$ | $2(\sigma_r - \sigma_\theta) + r\frac{d\sigma_r}{dr} = 0$ <br> $\dot{\varepsilon}_r = \frac{d}{dr}(r\dot{\varepsilon}_\theta)$ <br> $\dot{\varepsilon}_r = \frac{\dot{\sigma}_r - 2v\dot{\sigma}_\theta}{E}$ <br> $\dot{\varepsilon}_\theta = \frac{-v\dot{\sigma}_r + (1-v)\dot{\sigma}_\theta}{E}$ <br> $\sigma_r = p \ (r=a); \sigma_r = p_0 \ (r=b)$ |
| Formulation C | $\sigma_r, \sigma_\theta$ | $2(\sigma_r - \sigma_\theta) + r\frac{d\sigma_r}{dr} = 0$ <br> $\sigma_r - \sigma_\theta = r\frac{d\sigma_\theta}{dr}$ <br> $\sigma_r = p \ (r=a); \sigma_r = p_0 \ (r=b)$ |

These three formulations are embedded in three different PiNet-based solvers with the same network configurations to solve the same elastic cavity expansion problem; Table 2 presents the corresponding MSE of the predicted stress components ($\sigma_r$ and $\sigma_\theta$). When the governing equations are expressed in formulations **A** and **B**, PiNet shows poor performance with large MSE. A sharp decrease in MSE is observed when formulation **C** is embedded in the loss functions. This result reveals that the prediction accuracy of the PiNet is highly sensitive to the specific formulation of the governing equation used in the loss function. Thus, encoding the parsimony principle into the loss function by simplifying the governing equations improves the accuracy of PiNet-based solvers. This finding aligns with evidence in the literature that incorporating multiple constraints into the loss function can promote erroneous predictions or even convergence failure of PiNets (Wang *et al.*, 2021). Therefore, the development and deployment of robust PiNet-based solvers require careful optimization of the adopted structure of the loss function.

**Table 2 MSE of PiNet predictions using different formulations of the governing equations**

| Stress component | MSE | | |
|---|---|---|---|
| | formulation **A** | formulation **B** | formulation **C** |
| $\sigma_r$ | 6.7409 | 6.0032 | $3.19 \times 10^{-6}$ |
| $\sigma_\theta$ | 5.4717 | 4.4566 | $1.20 \times 10^{-6}$ |

Notably, whilst the simplified formulation **C** governing equations do not explicitly encompass all compatibility equations and constitutive relations, they still implicitly capture the underlying physics and serve as essential constraints for the PiNet output.

To further explore the reason why the parsimonious (formulation **C**) PiNet can outperform the regular (formulation **A** and **B**) PiNet, an analysis was conducted on the distribution of back-propagated gradients of loss functions with respect to the neural network parameters during training. Since the core of optimization during PiNet training is gradient descent, we track the gradients of loss based on the aforementioned three governing equation formulations with respect to weights parameter in each hidden layer. While PiNets here adopted for three governing equation formulations all comprise three hidden layers, the histogram of the back-propagated gradients for each hidden layer at the end of training is depicted in Fig. 3.

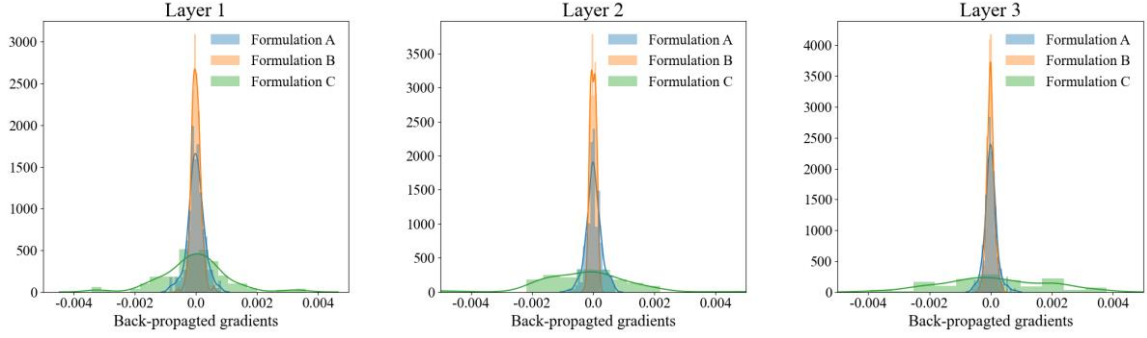

Fig. 3 Histograms of back-propagated gradients for three governing equation formulations embedded PiNets in each layer at the end of the training

Fig. 3 showcases that for both formulation **A** and **B** embedded PiNets, the back-propagated gradients of loss in each layer are sharply concentrated around zero. Whilst the back-propagated gradients of loss of the parsimonious (formulation **C**) PiNet distributed uniformly and overall attained significantly larger values. This implies that the adoption of the parsimony principle for loss functions can enlarge the back-propagated gradients of loss, as too small gradients may lead to slower training or even vanishing gradient pathologies (Wang *et al.*, 2021).

Owing to its superior performance, the parsimonious (formulation **C**) PiNet is adopted for all cases henceforth. Accordingly, the compatibility equations and constitutive relations for the cross-anisotropic elastic scenario can also be recast into a single equation:

$$\frac{1+v'}{E'}\sigma_r - \frac{1+v}{E}\sigma_\theta = -\frac{v'}{E'} \cdot r\frac{d\sigma_r}{dr} + \frac{1-v}{E} \cdot r\frac{d\sigma_\theta}{dr} \qquad (23)$$

# PiNet-based ODE solutions

## Model set-up

Four different cases are solved by the proposed PiNet including (*i*) elastic solutions in an isotropic medium; (*ii*) elastic solutions in an anisotropic medium; (*iii*) elastic-perfectly plastic solutions for Tresca yield criterion; (*iv*) elastic-perfectly plastic solutions for Mohr-Coulomb yield criterion. Table 3 summarizes the model parameters for these four cases.

Table 3 Variables adopted for the present cavity expansion problems

| Variables | Elastic case | | Elasto-plastic case | |
|---|---|---|---|---|
| | Case *i* | Case *ii* | Case *iii* | Case *iv* |
| $a$ (m) | 0.4 | 0.4 | 0.2 | 0.2 |
| $b$ (m) | 2.0 | 2.0 | 2.0 | 2.0 |
| $c$ (m) | - | - | 0.8 | 0.8 |
| $p$ (kPa) | 5.0 | 5.0 | - | - |
| $p_0$ (kPa) | 0 | 0 | 0 | 0 |
| $E$ (kPa) | $1.0 \times 10^5$ | $1.0 \times 10^5$ | $1.0 \times 10^5$ | $1.0 \times 10^5$ |
| $E'$ (kPa) | - | $0.8 \times 10^5$ | - | - |
| $v$ | 0.3 | 0.3 | 0.3 | 0.3 |
| $v'$ | - | 0.24 | - | - |
| $\phi$ (°) | - | - | - | 15 |
| $C$ (kPa) | - | - | - | 2.5 |
| $\sigma_Y$ (kPa) | - | - | 6.0 | - |

Since the only difference between these cases primarily stems from different governing equations and boundary conditions, a consistent neural network architecture is adopted for all cases. This approach ensures that the code requires only minimal modifications to replace loss functions when applied to different cases. A fully connected DNN is adopted with an input layer comprising only the radial distance $r$, three hidden layers of 16 neurons each, and an output layer composed of the radial stress $\sigma_r$ and tangential stress $\sigma_\theta$. This architecture is designed to strike a balance between accuracy and computational cost. The hyperbolic tangent activation function *tanh*(x) is adopted for its differentiability. The full-batch optimization algorithm and L-BFGS (Liu and Nocedal, 1989) optimizer are used for back-propagation to ensure fast convergence to the global minimum. An optimal learning rate was found to be 0.001 for this problem. All details of the PiNet configuration are summarized in Table 4.

Table 4 Details of the PiNet configuration for solving cavity expansion problems

|  | Value |
|---|---|
| Baseline network | DNN |
| Architecture | 1-16-16-16-2 |
| Activation function | $tanh(x)$ |
| Optimizer | L-BFGS |
| Batch size | 1 |
| Learning rate | 0.001 |

For cases *i* and *ii*, two separate PiNets with the same architecture shown in Table 4 are trained for the isotropic and anisotropic elastic problems. To calculate the residual value of different loss terms in the loss function, 50 collocation points were sampled in the computational domain using a uniform sampling method in addition to two boundary points: $r = a$ and $r = b$. Given the radius *r*, the stresses can be predicted using the PiNet nonlinear mapping function $\mathcal{NN}$ as follows:

$$[\sigma_r, \sigma_\theta] = \mathcal{NN}(r; \mathbf{W}, \mathbf{b}) \qquad (24)$$

For elasto-plastic cavity expansion problems, two separate neural networks are designed to sequentially predict the stress distribution in the elastic region and plastic region. This sequential design is necessary because the elastic stresses at the EP boundary ($r = c$) should be first calculated for use as boundary conditions to solve the ODEs in the plastic region. This process is depicted in Fig. 4.

When training the PiNet for the elastic region, 50 collocation points are uniformly sampled in the elastic region and two points on the boundary ($r = c$ and $r = b$) are selected. Another 50 collocation points are uniformly sampled for the plastic region and one point on the EP boundary ($r = c$) is selected. It is important to note that, unlike the algorithm employed for the elastic cases (cases *i* and *ii*), the internal pressure $p$ is not required to be a known variable in the elasto-plastic cases (cases *iii* and *iv*) given the problem is already well defined.

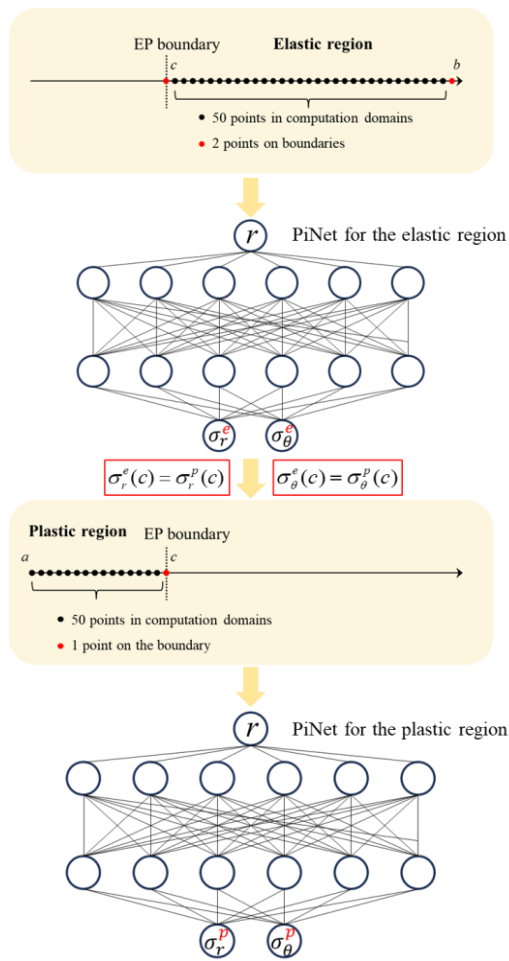

**Fig. 4 The process of using two separate PiNets to solve the elasto-plastic cavity expansion problem**

**Loss functions for each study case**

The simplified formulation **C** governing equations are combined with the aforementioned boundary conditions to arrive at the parsimonious loss functions for cases *i* and *ii* using the MSE operator:

$$\text{case i: } \mathcal{L} = \frac{1}{n_{col}} \left\| (\sigma_r(r) - \sigma_\theta(r)) - r\frac{d\sigma_\theta(r)}{dr} \right\|^2 + \frac{1}{n_{col}} \left\| 2(\sigma_r(r) - \sigma_\theta(r)) + r\frac{d\sigma_r(r)}{dr} \right\|^2$$

$$+ \frac{1}{n_b} \|\sigma_r(a) - p\|^2 + \frac{1}{n_b} \|\sigma_r(b) - p_0\|^2 \tag{25}$$

$$\text{case ii: } \mathcal{L} = \frac{1}{n_{col}} \left\| \frac{1+v'}{E'}\sigma_r(r) - \frac{1+v}{E}\sigma_\theta(r) - \left(-\frac{v'}{E'} \cdot r\frac{d\sigma_r(r)}{dr} + \frac{1-v}{E} \cdot r\frac{d\sigma_\theta(r)}{dr}\right) \right\|^2$$

$$+ \frac{1}{n_{col}} \left\| 2(\sigma_r(r) - \sigma_\theta(r)) + r\frac{d\sigma_r(r)}{dr} \right\|^2$$

$$+ \frac{1}{n_b} \|\sigma_r(a) - p\|^2 + \frac{1}{n_b} \|\sigma_r(b) - p_0\|^2 \tag{26}$$

Similarly, the parsimonious loss functions for cases *iii* and *iv* in the elastic region and plastic region can also be derived using the MSE operator:

$$\text{elastic region: } \mathcal{L} = \frac{1}{n_{col}} \left\| (\sigma_r^e(r) - \sigma_\theta^e(r)) - r\frac{d\sigma_\theta^e(r)}{dr} \right\|^2 + \frac{1}{n_{col}} \left\| 2(\sigma_r^e(r) - \sigma_\theta^e(r)) + r\frac{d\sigma_r^e(r)}{dr} \right\|^2$$

$$+ \frac{1}{n_b} \|(\sigma_r^e(c) - \alpha\sigma_\theta^e(c)) - \sigma_Y\|^2 + \frac{1}{n_b} \|\sigma_r^e(b) - p_0\|^2 \tag{27}$$

$$\text{plastic region: } \mathcal{L} = \frac{1}{n_{col}} \left\| 2(\sigma_r^p(r) - \sigma_\theta^p(r)) + r\frac{d\sigma_r^p(r)}{dr} \right\|^2 + \frac{1}{n_{col}} \|(\sigma_r^p(r) - \alpha\sigma_\theta^p(r)) - \sigma_Y\|^2$$

$$+ \frac{1}{n_b} \|\sigma_r^p(c) - \sigma_r^e(c)\|^2 + \frac{1}{n_b} \|\sigma_\theta^p(c) - \sigma_\theta^e(c)\|^2 \tag{28}$$

It should be noted that the loss functions for both cases *iii* and *iv* can be expressed by unified equations but the values of $\alpha$ and $\sigma_Y$ are determined by the adopted yield criterion. All unknown variables are either the direct output of the PiNet or calculated by AD.

**Solutions for elastic ODEs**

The evolution of loss values during PiNet training for cases *i* and *ii* is shown in Fig. 5. For both cases, there is an initial steep decline in the loss, followed by a more gradual decreasing trend during training. The PiNets approach convergence (plateau in the loss) at approximately 300 epochs with negligible change beyond 400 epochs. Importantly, the loss for both cases converges to values below $10^{-4}$, indicating successful attainment of the global minimum.

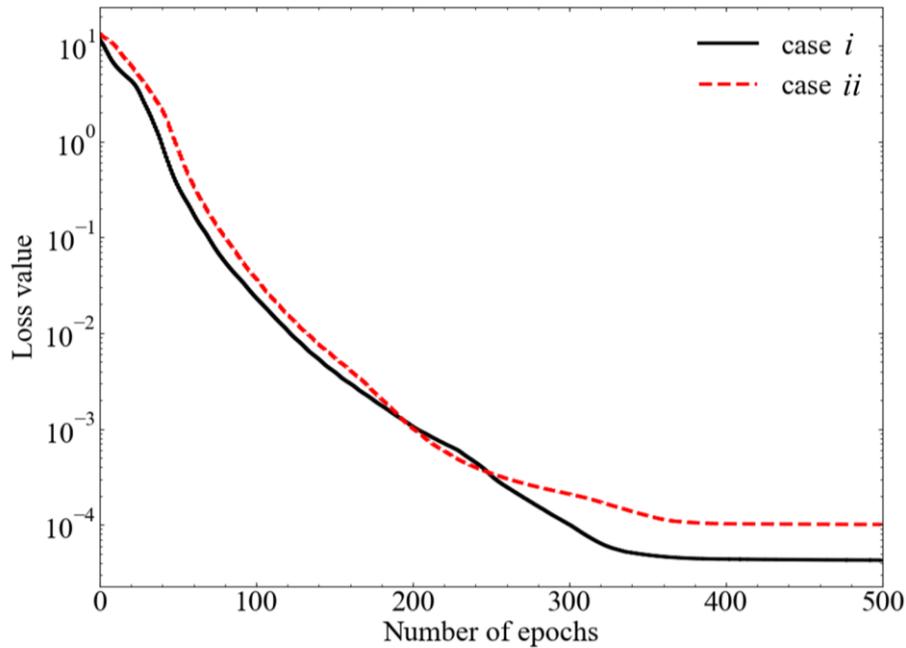

**Fig. 5 Evolution of the loss during PiNet training for the elastic cavity expansion problems (cases *i* and *ii*)**

The predicted stress distributions generated by the trained PiNets are compared with the exact stress distributions (Yu, 2000) in Fig. 6 and Fig. 7 for case *i* and case *ii*, respectively. In both cases, the predicted radial and tangential stress distributions exhibit excellent agreement with the exact values.

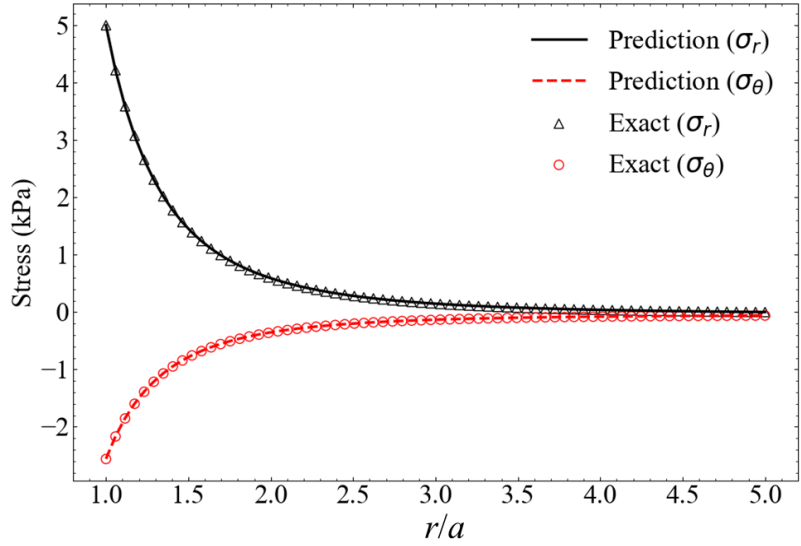

**Fig. 6 Comparison between PiNet-predicted radial and tangential stress fields and the exact theoretical stress distributions for case *i***

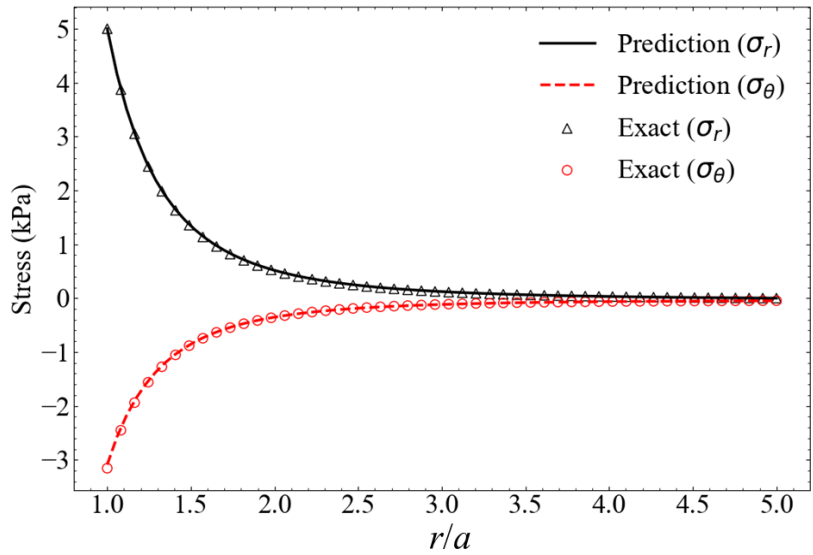

**Fig. 7 Comparison between PiNet-predicted radial and tangential stress fields and the exact theoretical stress distributions for case *ii***

In Fig. 8 the point-wise errors between the predicted and exact stress distributions for both elastic cases are presented. Generally, the point-wise errors are rather small and are within an order of magnitude of $\mathcal{O}(10^{-2})$, showing an acceptable PiNet approximation of the stress fields. A comparative analysis of the point-wise errors between case *i* and case *ii* reveals that the errors observed in case *i* are generally lower compared to those in case *ii*. This phenomenon can be attributed to the higher complexity associated with anisotropy, which introduces additional challenges in generating accurate predictions compared to the relatively simpler isotropic cases.

Additionally, in both cases, relatively larger errors can be observed for the tangential stresses compared with the radial stresses, particularly near the inner boundary. This is because the radial stress boundary values are explicitly incorporated into the loss function as boundary constraints during PiNet training (Eq. 25, Eq. 26), whereas tangential stress boundary values are not considered as prior information. Consequently, the tangential stresses are constrained only by the governing equations while the radial stresses benefit from the simultaneous enforcement of both governing equations and boundary conditions. Moreover, the point-wise error near the inner boundary is slightly larger than that near the outer boundary owing to the greater magnitudes at the inner boundary. Nevertheless, the MSE and $R^2$ values determined for the whole computation show that the proposed PiNets provide highly accurate solutions for the elastic cavity expansion problem (see Table 5).

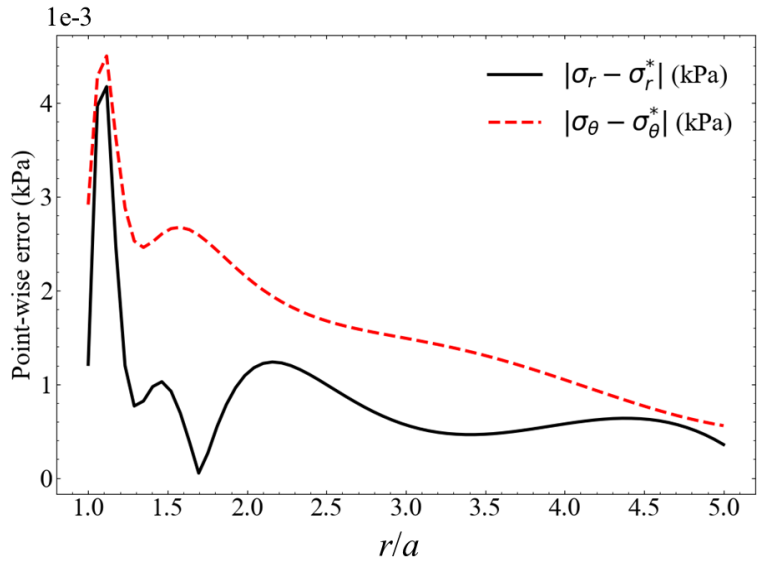

(a)

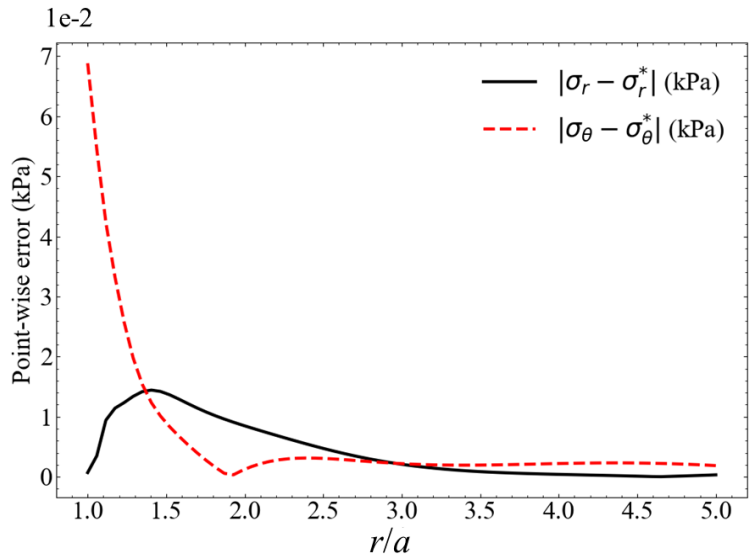

(b)

**Fig. 8 The point-wise error of the PiNet-predicted elastic stress distributions in both radial and tangential directions: (a) case *i*; (b) case *ii***

**Table 5 MSE and $R^2$ values for the PiNets solutions of the elastic cavity problems**

| Accuracy | Case *i* | | Case *ii* | |
| --- | --- | --- | --- | --- |
| | Radial stress | Tangential stress | Radial stress | Tangential stress |
| MSE | $3.19 \times 10^{-6}$ | $1.20 \times 10^{-6}$ | $4.53 \times 10^{-5}$ | $2.0 \times 10^{-4}$ |
| $R^2$ | 1.0000 | 1.0000 | 1.0000 | 0.9996 |

**Solutions for elasto-plastic ODEs**

The evolution of loss values during PiNet training for the elasto-plastic cavity expansion problem using Tresca and Mohr-Coulomb criteria (cases *iii* and *iv*) are shown in Fig. 9. Again, there is an initial rapid decline in the loss values, followed by a gradual convergence to a stable and small value. While the plateau beyond 600 epochs indicates that the PiNets have successfully achieved accurate solutions for the elasto-plastic cavity expansion problem. There are some notable differences between the cases. First, the loss relating to the plastic region is larger than that in the elastic region for both case *iii* and case *iv*, implying the challenge for PiNet in capturing plastic behaviours. Second, the loss for the plastic region governed by the Tresca criterion (case *iii*) shows faster convergence than that for the Mohr-Coulomb criterion (case *iv*), highlighting some sensitivity of the PiNet solutions to the adopted yield criterion.

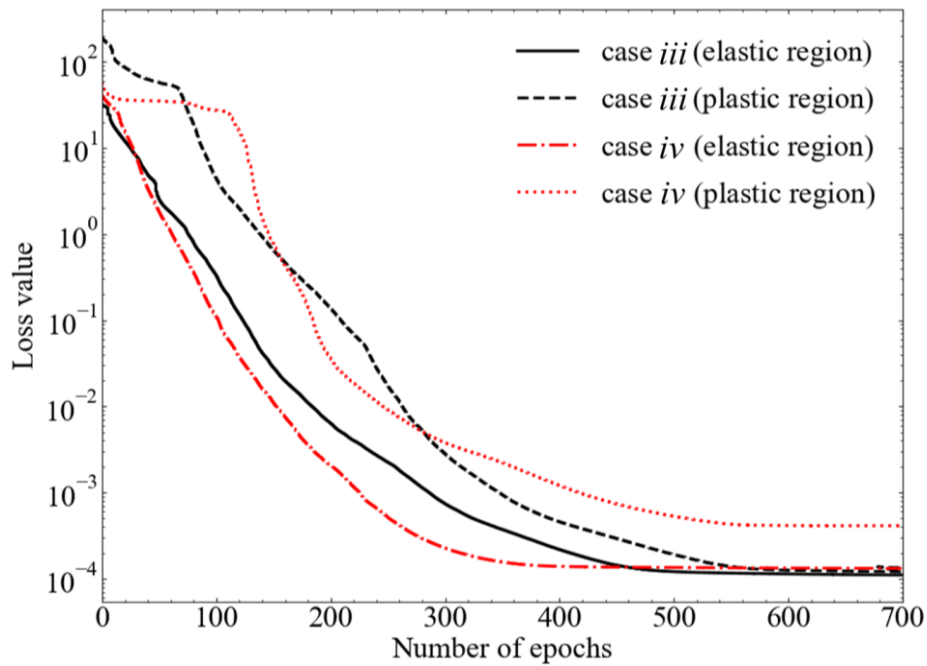

**Fig. 9 Evolution of the loss during PiNet training for the elasto-plastic cavity expansion problems (cases *iii* and *iv*)**

The PiNet-predicted stress distribution is compared with the exact theoretical values for both cases *iii* and iv in Fig. 10 and Fig. 11, respectively. It can be seen that the trained PiNet can predict the soil stress fields surrounding an expanded cavity accurately in the context of elastic-perfectly plastic theory. Despite the internal pressure *p* is not considered a known boundary condition in cases *iii* and *iv*, the PiNet can still identify the inner pressure with high accuracy. However, due to the focus on stress analysis in this research, only the relationship between the inner pressure *p* and the plastic zone radius *c* could be captured. The calculation of the relationship between displacement and the plastic zone radius was not conducted, which represents an inherent limitation of the proposed method.

The point-wise prediction error for both cases is presented in Fig. 12. In both cases, errors are acceptably small and within an order of magnitude of $\mathcal{O}(10^{-2})$. The errors within the plastic region are notably larger and noisier than within the elastic region, suggesting that accurately capturing plastic behaviours using PiNet poses a greater challenge. Again, the point-wise errors are smaller when using the Tresca criterion compared with the Mohr-Coulomb criterion implying the influence of the yield criterion on the ability of PiNet to capture and predict these mechanical behaviours.

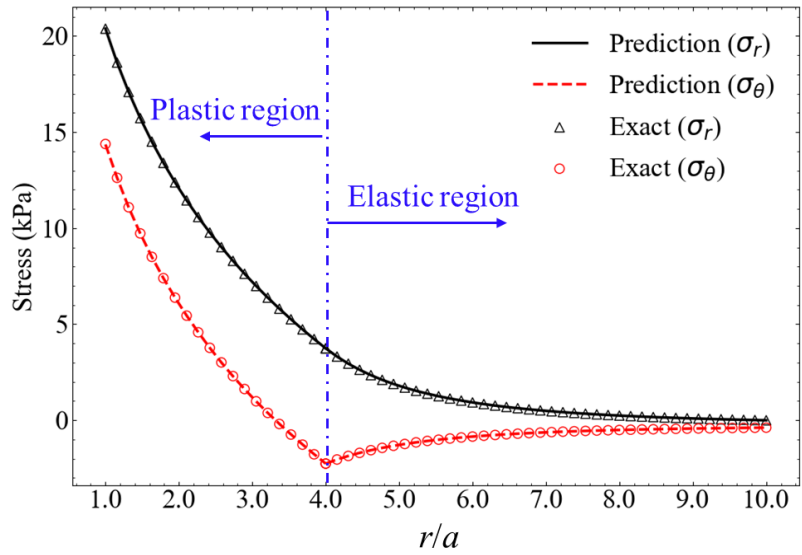

Fig. 10 Comparison between PiNet-predicted radial and tangential stress fields and the exact theoretical stress distributions for case *iii*

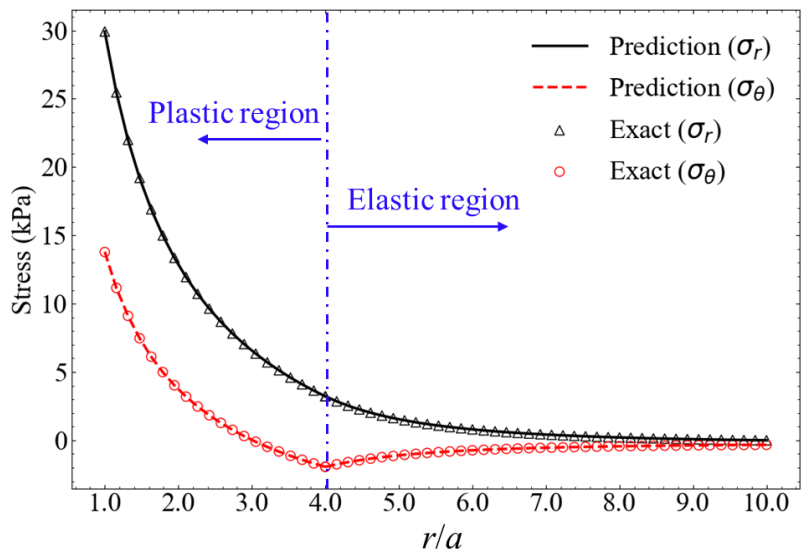

Fig. 11 Comparison between PiNet-predicted radial and tangential stress fields and the exact theoretical stress distributions for case *iv*

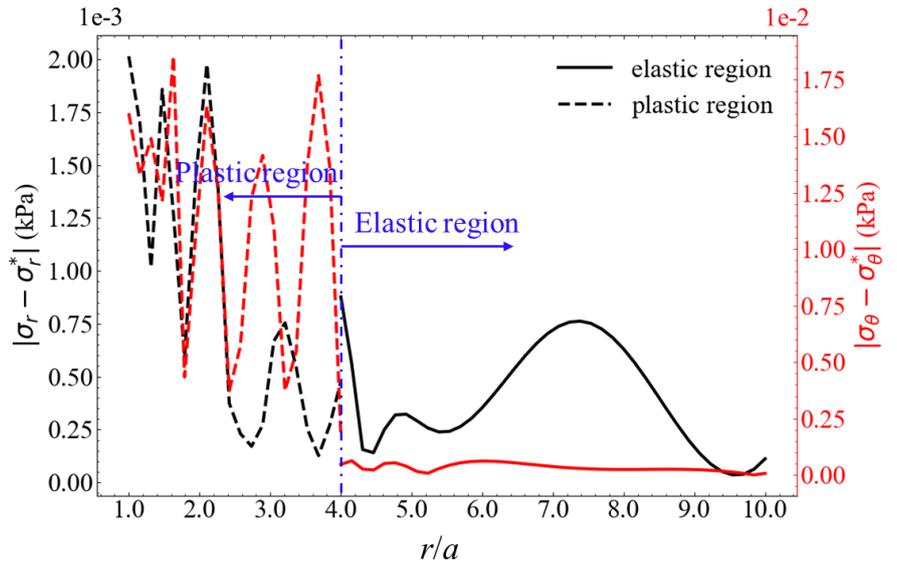

(a)

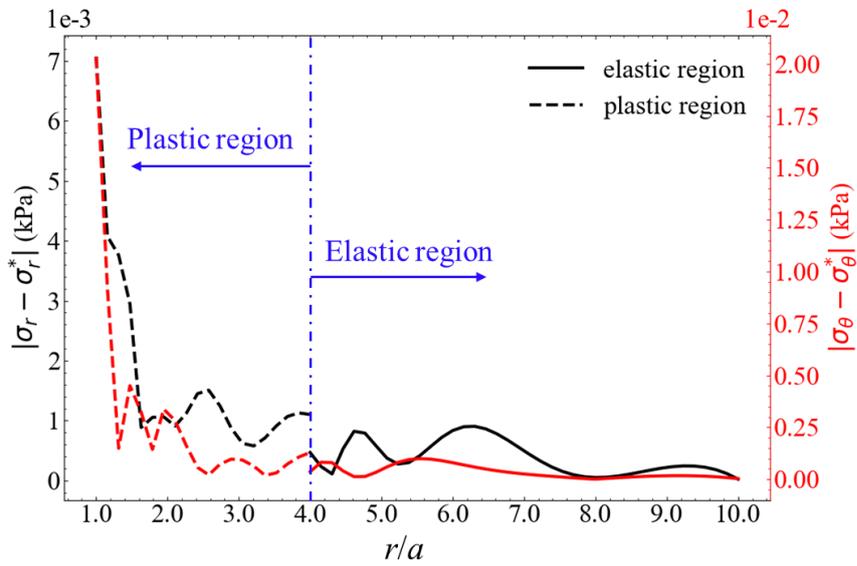

(b)

**Fig. 12 The point-wise error of the PiNet-predicted elasto-plastic stress distributions in both radial and tangential directions: (a) case *iii*; (b) case *iv***

The MSE and $R^2$ values in the plastic and elastic regions for case *iii* and case *iv* are shown in Table 6. The MSE for these cases are all consistently small, and the values of $R^2$ are all close to 1.0. These results suggest that the proposed PiNet solver can effectively capture the elasto-plastic behaviours of an expanded cavity, demonstrating their potential for accurate and reliable meshless modelling of geotechnical problems.

Table 6 MSE and $R^2$ values for the PiNets solutions of the elasto-plastic cavity problems

| Accuracy | Case iii | | | |
|---|---|---|---|---|
| | Elastic region | | Plastic region | |
| | Radial stress | Tangential stress | Radial stress | Tangential stress |
| MSE | $1.97\times10^{-7}$ | $1.38\times10^{-7}$ | $3.17\times10^{-7}$ | $2.0\times10^{-4}$ |
| $R^2$ | 1.0000 | 1.0000 | 1.0000 | 1.0000 |

| Accuracy | Case iv | | | |
|---|---|---|---|---|
| | Elastic region | | Plastic region | |
| | Radial stress | Tangential stress | Radial stress | Tangential stress |
| MSE | $4.45\times10^{-7}$ | $6.55\times10^{-7}$ | $1.48\times10^{-5}$ | $2.0\times10^{-4}$ |
| $R^2$ | 1.0000 | 1.0000 | 1.0000 | 1.0000 |

## Conclusions

This paper has introduced a novel parsimonious prior information-based neural network (PiNet), as a universal function approximator for solving elastic and elasto-plastic engineering problems. Unlike conventional deep neural networks that heavily rely on large amounts of labelled data, the solution of partial differential equations (PDEs) or ordinary differential equations (ODEs) can be obtained by these PiNets without any labelled data. The effectiveness of this approach was verified through its application to one-dimensional cavity expansion problems with increasing complexity.

To encode prior knowledge, constitutive relations, equilibrium equations, compatibility equations and boundary conditions were incorporated into the loss function of the PiNets. These governing equations can be formulated using different mathematical equations with different levels of complexity for incorporation into the PiNet loss functions. A comparative study was conducted to investigate the sensitivity of the accuracy of the PiNet solutions to these formulations. The results demonstrated that employing parsimonious loss functions, by simplifying the governing equations and reducing the output dimensions, significantly enhances the accuracy of the PiNet solutions. Incorporating multiple governing equations within the loss functions was observed to introduce potential inaccuracies and spurious predictions. Moreover, by analysing the back-propagated gradients of loss functions with respect to neural network weight parameters, the parsimonious loss functions can achieve larger back-propagated gradient values, contributing to fast convergence. This property ensures that the optimization process efficiently updates the network weights towards the optimal solution. Thus, the present parsimonious loss function structure is recommended to strike a balance between model simplicity and the ability to capture the essential mechanical responses. This approach not only facilitates implementation but also enhances the accuracy and reliability of the results.

Capturing plastic behaviour using PiNets is notably more challenging than capturing elastic behaviour. By using parsimonious PiNets for solving elasto-plastic problems, taking both Tresca and Mohr-Coulomb criteria into account, the corresponding elasto-plastic solutions showed excellent agreement with the exact solutions.

The proposed research framework is generic for simulating various geotechnical problems with known prior information and physics laws. Even if the underlying physical laws are not fully understood, a hybrid data-driven and physics-driven strategy can be adopted to simulate geotechnical problems using a very limited amount of data.

## Declaration of competing interest

The authors declare that they have no known competing financial interests or personal relationships that could have appeared to influence the work reported in this paper.

## Data Availability Statement

All data that support the findings of this study are available from the corresponding author upon reasonable request.

## Acknowledgements

This research was financially supported by the Research Grants Council (RGC) of Hong Kong Special Administrative Region Government (HKSARG) of China (Grant No.: 15220221). The second author is supported by the Royal Society under the Newton International Fellowship.

## Notation

| | |
|---|---|
| $a, b$ | inner and outer radius |
| $\alpha$ | material parameter |
| $\boldsymbol{b}$ | bias vector |
| $\mathcal{B}_D(t), \mathcal{B}_N(t)$ | prescribed functions on the Dirichlet and Neumann boundaries |
| $c$ | plastic zone radius |
| $C$ | cohesion of the soil |
| $\mathcal{D}[\cdot]$ | nonlinear differentiation operator |
| $E'$, $E$ | Young's modulus in the $r$ direction and in the isotropic surface, respectively |
| $\varepsilon_r$, $\varepsilon_\theta$ | principle strains in the radial and tangential directions, respectively |
| $\mathcal{I}_0(x)$, $\mathcal{I}_1(x)$ | initial states of the system |
| $L$ | total number of the deep neural network (DNN) hidden layers |
| $n_{col}$ | number of collocation points in the computational domain |
| $n_b$ | number of collocation points on the boundaries |
| $n_i$ | number of collocation points at initials |
| $n_d$ | number of data points |
| $\mathbf{n}$ | unit outward normal vector of the boundaries |
| $\mathcal{NN}$ | nonlinear mapping function |
| $p$, $p_0$ | internal and external pressure |
| $r$ | radial distance |
| $(r, \theta)$ | polar coordinate system |
| $s_u$ | undrained strength of the soil |
| $t$, $\boldsymbol{x}$ | temporal and spatial coordinates |
| $T$ | maximum computation time |
| $u$ | radial displacement |
| $v'$, $v$ | Poisson's ratio in the $r$ direction and in the isotropic plane, respectively |
| $w_1$ | weight assigned to the $\mathcal{L}_{PDE}$ loss term |
| $w_2$ | weight assigned to the $\mathcal{L}_{BC}$ loss term |
| $w_3$ | weight assigned to the $\mathcal{L}_{IC}$ loss term |
| $w_4$ | weight assigned to the $\mathcal{L}_{data}$ loss term |
| $\mathbf{W}$ | weight matrix |
| $\mathbf{z}_0$, $\mathbf{z}_{L+1}$ | input and output tensors of the DNN |
| $\mathbf{z}_{l-1}$, $\mathbf{z}_l$ | input and output tensors of the $l^{th}$ layer |
| $\partial\Omega_D$, $\partial\Omega_N$ | Dirichlet and Neumann boundaries |
| $\sigma_r$, $\sigma_\theta$ | principle stresses in the radial and tangential directions, respectively |
| $\sigma_r^e$, $\sigma_\theta^e$, $\sigma_r^p$, $\sigma_\theta^p$ | radial and tangential stresses in the elastic and plastic regions, respectively |
| $\sigma_Y$ | the soil yield stress |
| $\sigma$ | activation function |
| $\phi$ | friction angle of the soil |
| ∇ | gradient operator |
| Ω | domain of input variables |